\documentclass[a4paper,BCOR=10mm]{scrbook}
\usepackage{amsmath,bm}
\usepackage{amsfonts}
\usepackage{amssymb}
\usepackage{epsfig}
\usepackage{graphicx}
\usepackage{color}
\usepackage{natbib}

\definecolor{red}{rgb}{.98,0,0}

\newcommand{\sgn}{\mathrm{sgn}}

\begin{document}

\chapter[Self-organized criticality in neural network models]{Self-organized criticality \newline in neural network models}

\hspace*{1.75em}
{\textit{Matthias Rybarsch and Stefan Bornholdt}}

\vspace*{1mm}

{\noindent \hspace*{1.8em} \small{Institut f\"{u}r Theoretische Physik, Universit\"at Bremen, D-28359 Bremen, Germany}}

\let\thefootnote\relax\footnote{In ``Criticality in Neural Systems'', Niebur E, Plenz D, Schuster HG (eds.) 2013 (in press).}

\vspace*{5mm}
\section{Introduction}
Information processing by a network of dynamical elements is a delicate matter: 
Ava\-lanches of activity can die out if the network is not connected enough or 
if the elements are not sensitive enough; on the other hand, 
activity avalanches can grow and spread over the entire network 
and override information processing as observed in epilepsy. 

Therefore, it has long been argued that neural networks have to 
establish and maintain a certain intermediate level of activity 
in order to keep away from the regimes of chaos and silence \citep{Langton:1990,Herz:1995,Bak:2001,Bornholdt:2003}. 
Similar ideas were also formulated in the context of genetic networks 
where Kauffman postulated that information processing
in these evolved biochemical networks would be optimal near 
the ``edge of chaos'', or criticality, of the dynamical percolation 
transition of such networks \citep{Kauffman:1993}.

In the wake of self-organized criticality (SOC), it was asked if also 
neural systems were self-organized to some form of criticality 
\citep{Bak:1988}. 
In addition, actual observations of neural oscillations within the human 
brain were related to a possible SOC phenomenon \citep{Linkenkaer-Hansen:2001}.
An early example of a SOC model that had been adapted to be 
applicable to neural networks is the model by  \citet{Eurich:2002}.
They show that their model of the random neighbor Olami-Feder-Christensen 
universality class exhibits (subject to one critical coupling parameter) 
distributions of avalanche sizes and durations which 
they postulate could also occur in neural systems. 

Another early example of a model for self-organized critical neural networks 
\citep{Bornholdt:2001,Bornholdt:2003} drew on an alternative  
approach to self-organized criticality based in dynamical networks 
\citep{Bornholdt:2000a}. 
Here networks are able to self-regulate towards and maintain a critical system state, 
via simple local rewiring rules which are plausible in the biological context. 

After these first model approaches, indeed strong evidence for criticality 
in neural systems has been found in terms of spatio-temporal activity avalanches, 
first in the seminal work of \citet{Beggs:2003}. 
Much further experimental evidence has been found since, 
which we will briefly review below. These experimental findings sparked 
intense research on dynamical models for criticality and avalanche dynamics 
in neural networks, which we also give a brief overview below. 
While most models emphasized biological and neurophysiological detail, 
our path here is different:

The purpose of this review is to pick up the thread of the early 
self-organized critical neural network model \citep{Bornholdt:2003} 
and test its applicability in the light of experimental data. 
We would like to keep the simplicity of the first spin model in the light 
of statistical physics, while lifting the drawback of a spin formulation
w.r.t. the biological system \citep{Rybarsch:2012a}. 
We will study an improved model and show that it adapts to criticality 
exhibiting avalanche statistics that compare well with experimental data 
without the need for parameter tuning \citep{Rybarsch:2012b}.

\section{Avalanche dynamics in neuronal systems}
\subsection{Experimental results}
Let us first briefly review the experimental studies on neuronal
avalanche dynamics.  In 2003, Beggs and Plenz published their findings
about a novel mode of activity in neocortical neuron circuits
\citep{Beggs:2003}. During \emph{in-vitro} experiments with cortex
slice cultures of the rat where neuronal activity in terms of local
field potentials was analyzed via a 8x8 multi-electrode array, they
found evidence of spontaneous bursts and avalanche-like propagation of
activity followed by silent periods of various lengths. The observed
power-law distribution of event sizes indicates that the neuronal
network is maintained in a critical state. In addition, they found
that the precise spatio-temporal patterns of the avalanches are stable
over many hours and also robust against external perturbations
\citep{Beggs:2004}. They conclude that these neuronal avalanches might
play a central role for brain functions like information storage and
processing. Also during developmental stages of \emph{in-vitro} cortex
slice cultures from newborn rats, neuronal avalanches were found,
indicating a homeostatic evolution of the cortical network during
postnatal maturation \citep{Stewart:2007}. Moreover, also cultures of
dissociated neurons were found to exhibit this type of spontaneuous
activity bursts in different kinds of networks, like rat hippocampal
neurons and leech ganglia \citep{Mazzoni:2007}, as well as dissociated
neurons from rat embryos \citep{Pasquale:2008}.

Aside from these \emph{in-vitro} experiments, extensive studies \emph{in-vivo} have since been conducted. The emergence of spontaneous neuronal avalanches has been shown in anaesthesized rats during cortical development \citep{Gireesh:2008} as well as in awake rhesus monkeys during ongoing cortical synchronization \citep{Petermann:2009}.

The biological relevance of the avalanche-like propagation of activity in conjunction with a critical state of the neuronal network has been emphasized in several works recently. Such network activity has proven to be optimal for maximum dynamical range \citep{Kinouchi:2006,Shew:2009,Larremore:2011}, maximal information capacity and transmission capability \citep{Shew:2011} as well as for a maximal variability of phase synchronization \citep{Yang:2012}.

\subsection{Existing models}
The experimental studies with their rich phenomenology of spatio-temporal patterns
sparked a large number of theoretical studies and models 
for criticality and self-organization in neural networks, ranging from very simple toy models to detailed representations of biological functions. Most of them try to capture self-organized behavior with emerging avalanche activity patterns, with scaling properties similar to the experimental power-law event size or duration distributions. 

As briefly mentioned above, early works as \citet{Bornholdt:2000a} and \citet{Bornholdt:2001, Bornholdt:2003} focus on simple mechanisms for self-organized critical dynamics in spin networks, which also have been discussed in a wider context \citep{Gross:2008}. 
These models represent an approach aiming at utmost simplicity of the model, 
quite similar to the universality viewpoint of statistical mechanics, 
rather than faithful representations of neurobiological and biochemical detail.
Nevertheless they are able to self-regulate towards and maintain a critical system state, 
manifested in features as a certain limit cycle scaling behavior, 
via simple local rewiring rules which are still plausible in the biological context. 
We will have a closer look at these models in the following section \ref{sec:Models}, 
because they provide some of the groundwork for current models.

Regarding neuronal avalanches, a 2002 work by Eurich et al.\ investigates networks of globally coupled threshold elements which are related to integrate-and-fire neurons. They present a model which, after proper parameter tuning, exhibits avalanche-like dynamics with distinctive distributions of avalanche sizes and durations as expected at a critical system state \citep{Eurich:2002}.

It is notable that these models came up even before experimental evidence was found for the existence of neuronal avalanches by \citet{Beggs:2003}. Understandably, extensive studies have been done on avalanche models following their discovery. Again, most models have their mechanisms of self-organization motivated by brain plasticity.

A 2006 model proposed by de Arcangelis et al.\ consists of a model electrical network on a square lattice, where threshold firing dynamics, neuron refractory inhibition and activity-dependent plasticity of the synaptic couplings, represented by the conductance of the electrical links, serve as the basis for self-organization. Neuron states are characterized by electrical potentials, which may be emitted as action potentials to neighboring nodes once a certain threshold has been reached. With these incoming currents, the neighbor sites can eventually also reach their activation threshold and thus activity avalanches can propagate through the network. Avalanches are triggered by external  currents to specific input sites. Following the activation of a node, the coupling conductances are increased by a small value for each link which has carried a firing current. On the other hand, after completing a whole avalanche, all couplings in the network have their conductance reduced by the average of the increase which has taken place before during the avalanche propagation. This way, those couplings which carry many signals, will effectively get stronger connections, while the rather inactive connections will be reduced and subsequently pruned from the network. Indeed, the model evolves to a critical state with power-law scaling of avalanche sizes \citep{Arcangelis:2006}. In a following work, the same behavior of such an adaptive model could also be observed on a class of scale-free networks (namely Apollonian networks), which is argued to be more appropriate as an approach to real neuronal networks than a lattice would be \citet{Pellegrini:2007}.

A related model (in terms of insertion of links or facilitation of weights where signals have been passed) has been proposed by \citet{Meisel:2009}. The authors focus on the interplay between activity-driven vs.\ spike-time-dependent plasticity and link their model to a phase transition in synchronization of the network dynamics. The temporal sequence of node activations serves as the key criterion for the topological updates. While they do not specifically discuss avalanche-like activity patterns, one observes power-law distributed quantities like correlation measures or synaptic weights in the self-organized states which point to dynamical criticality.

While the last three models mentioned above are set up to strengthen
those couplings over which signals have been transmitted, a kind of
opposite approach was proposed by \citet{Levina:2007}. In their model,
synaptic depression after propagation of activity over a link --
biologically motivated by the depletion of neuro-transmitter
ressources in the synapse -- is the key mechanism which drives their
model to a critical behavior. The network consists of fully connected
integrate-and-fire neurons whose states are described by a membrane
potential. This potential is increased by incoming signals from other
sites or by random external input, and may cause the site to emit
signals when the activation threshold is exceeded. Following such a
firing event, the membrane potential is reduced by the threshold
value. Again, a single neuron starting to fire after external input,
may set off an avalanche by passing its potential to other sites,
which in turn can exceed their activation threshold, and so on. The
couplings in this model take non-discrete values, directly related to
the biologically relevant amount of neuro-transmitter available in
each synapse. In short, whenever a signal is passed by a synapse, its
value will be decreased. The coupling strength is on the other hand
recovering slowly towards the maximum value in periods when no firing
events occur. The authors also extend the model to consider leaky
integrate-and-fire neurons, and find a robust self-organization
towards a critical state, again with the characteristic power-law
distribution of avalanche sizes. In a later work \citep{Levina:2009},
the authors further investigate the nature of the self-organization
process in their model and discuss the combination of first-\ and
second-order phase transitions with a self-organized critical phase. 

Meanwhile, field-theoretic approaches to fluctuation effects helped to
shed light on the universality classes to expect in critical neural
networks \citep{Buice:2007} and the presence of SOC in
non-conservative network models of leaky neurons were linked to the
existence of alternating states of high vs.\ low activity, so-called
up- and down-states \citep{Millman:2010}. It has been shown that
anti-Hebbian evolution is generally capable of creating a dynamically
critical network when the anti-Hebbian rule affects only symmetric
components of the connectivity matrix. The anti-symmetric component
remains as an independent degree of freedom and could be useful in
learning tasks \citep{Magnasco:2009}. Another model highlights the
importance of synaptic plasticity for a phase transition in general
and relates the existence of a broad critical regime to a hierarchical
modularity \citep{Rubinov:2011}. The biological plausibility of
activity-dependent synaptic plasticity for adaptive self-organized
critical networks has further been stressed
recently \citep{Droste:2012}. Also, robustness of critical brain
dynamics to complex network topologies has been
emphasized \citep{Larremore:2012}. 

The relevance of the critical state in neuronal networks for a brain
function as learning was underlined by \citet{Arcangelis:2010}, where
the authors find that the perfomance in learning logical rules as well
as time to learn are strongly connected to the strength of plastic
adaptation in their model, which at the same time is able to reproduce
critical avalanche activity. In a most recent
work \citep{Arcangelis:2012}, the same authors present a new
variant of their electrical network model. They again use facilitation
of active synapses as their primary driving force in the
self-organization, but now focus more on activity correlation between
those nodes which are actually active in consecutive time steps. Here,
they investigate the emergence of critical avalanche events on a
variety of different network types, as for example regular lattices,
scale-free networks, small-world networks or fully connected
networks. 

Also most recently, \citet{Lombardi:2012} investigate the temporal
organization of neuronal avalanches in real cortical networks. The
authors find evidence that the specific waiting time distribution
between avalanches is a consequence of the above-mentioned up- and
down-states, which in turn is closely linked to a balance of
excitation and inhibition in a critical network. 

\begin{figure}[h!tbp]
\centering
\includegraphics[width=0.80\textwidth]{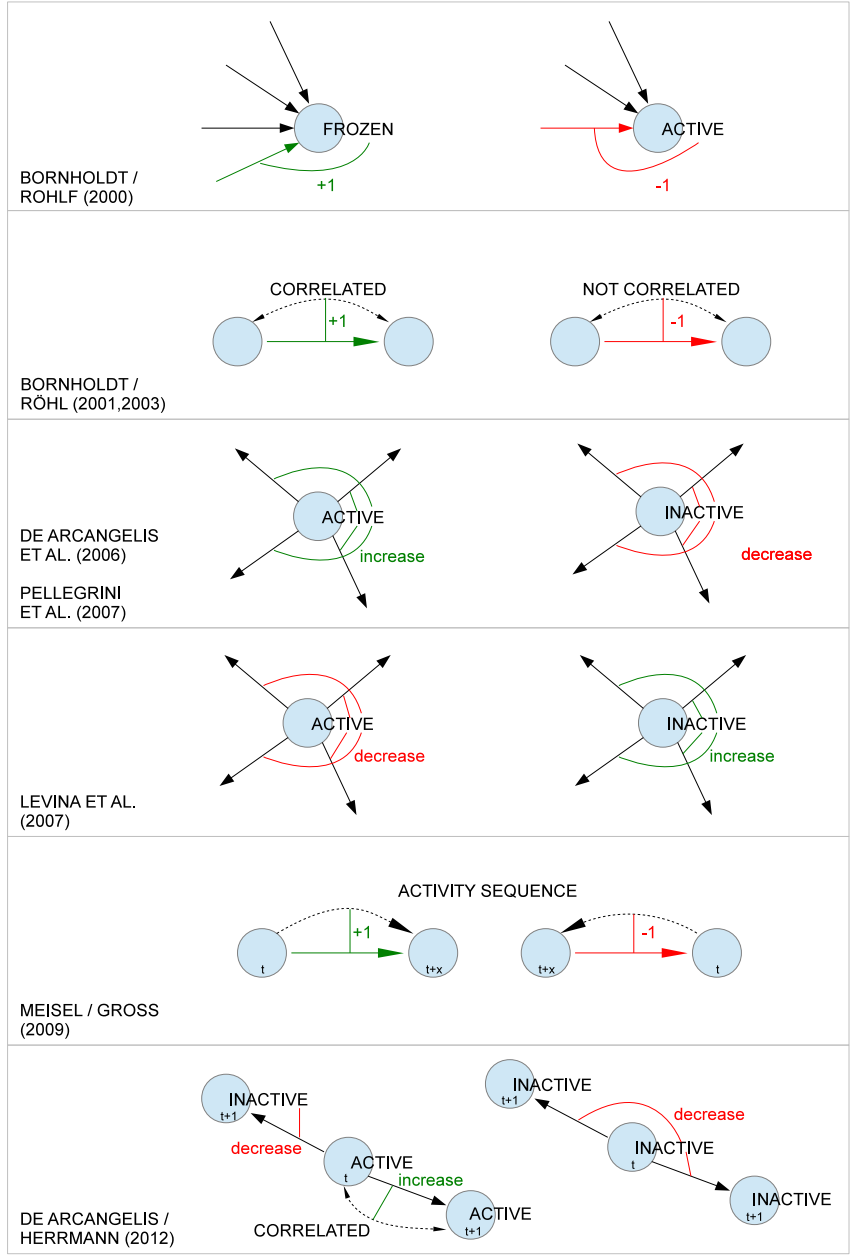}
\caption{\small Schematic illustration of some of the different approaches to self-organization in neural network models. Rows 1 - 2: Links are either added (denoted by +1; green link) or removed (denoted by -1; red link) as a function of node activity or correlation between nodes. Rows 3 - 4: Here, activity or inactivity of a node affects all outgoings links (thin lines). All weights of the outgoing links from a node are decreased (red) or increased (green) as a function of node activity. Row 5: Links are created and facilitated when nodes become active in the correct temporal sequence. Links directed against the sequence of activation are deleted.
Row 6: Positive correlation in the activity between two nodes selectively increases the corresponding link, whereas there is non-selective weight decrease for links between uncorrelated or inactive nodes.}
\label{fig:rewiring-models-schematic}
\end{figure}

While the proposed organization mechanisms strongly differ between the
individual models (see Figure~\ref{fig:rewiring-models-schematic} for
a cartoon representation of the various mechanisms), the resulting
evolved networks tend to be part of only a few fundamental
universality classes, exhibiting e.g.\ avalanche statistics in a
similar way as in the experimental data, as power-law distributions at
an exponent of $-3/2$. With the recent, more detailed models in mind,
we are especially interested in the underlying universality of
self-organization, also across other fields. Considering the enormous
interest in neuronal self-organization, we here come back to our older
spin models \citep{Bornholdt:2000a, Bornholdt:2003} and develop a
new basic mechanism in the light of better biological plausibility of
these models.

\section{Simple models for self-organized critical adaptive neural networks}\label{sec:Models}

\subsection{A first approach: node activity locally regulates connectivity}\label{sec:BornholdtRohlf}

In the very minimal model of a random threshold network \citep{Bornholdt:2000a} a simple local mechanism for topology evolution based on node activity has been defined, which is capable of driving the network towards a critical connectivity of $K_c = 2$. Consider a network composed of $N$ randomly and asymmetrically connected spins ($\sigma_i = \pm 1$), which are updated synchronously in discrete time steps via a threshold function of the inputs they receive:
\begin{equation}
	\sigma_i(t+1) = \sgn(f_i(t))
	\label{eq:rohlf-update}
\end{equation}
using
\begin{equation} 
\sgn(x)= 
\left\{ 
\begin{array}{c} 
+1, \ \ \  x\geq 0 \noindent \\  
-1, \ \ \ x < 0  
\end{array} 
\right. 
\end{equation}  
and 
\begin{equation}
	f_i(t) = \sum_{j=1}^N c_{ij} \sigma_j(t) + h .
\end{equation}
where the link weights have discrete values $c_{ij} = \pm 1$ (or $c_{ij} = 0$ if node $i$ does not receive input from node $j$). In the minimal model, activation thresholds are set to $h=0$ for all nodes. A network run is started with random initial configuration and iterated until either a fixed point attractor or a limit cycle attractor is reached.
The attractor of a network is where its dynamics ends up after a while, which is either a fixed point of the dynamics (all nodes reach a constant state) or a limit cycle of the whole network dynamics. A limit cycle in these discrete dynamical models is a cyclic sequence of a finite number of activation patterns. 

For the topological evolution, a random node $i$ is selected and its activity during the attractor period of the network is analyzed. The network is observed until such an attractor is reached; and afterwards, activity of the single node during that period is measured. 
In short, if node $i$ changes its state $\sigma_i$ at least once during the attractor, a random one of the existing non-zero in-links $c_{ij}$ to that node is removed. If, vice versa, $\sigma_i$ remains constant throughout the attractor period, a new non-zero in-link $c_{ij}$ from a random node $j$ is added to the network. 

In one specific among several possible realizations of an adaptation algorithm, the average activity $A(i)$ of node $i$ over an attractor period from $T_1$ to $T_2$ is defined as
\begin{equation}
	A(i) = \frac{1}{T_2-T_1} \sum_{t=T_1}^{T_2} \sigma_i (t).
\end{equation}
Topological evolution is now imposed in the following way:
\begin{enumerate}
	\item A random network with average connectivity $K_{ini}$ is created and node states are set to random values.
	\item Parallel updates according to \eqref{eq:rohlf-update} are executed until a previous state reappears (i.e. until a dynamical attractor is reached).
	\item Calculate $A(i)$ for a randomly selected node $i$. If $\vert A(i) \vert = 1$, node $i$ receives a new in-link of random weight $c_{ij} = \pm 1$ from a random other node $j$. Otherwise (i.e. if the state of node $i$ changes during the attractor period), one of the existing in-links is set to zero.
	\item Optional: a random non-zero link in the network has its sign reversed.
\end{enumerate}

\begin{figure}[h!tbp]
	\centering
	\includegraphics[width=85mm]{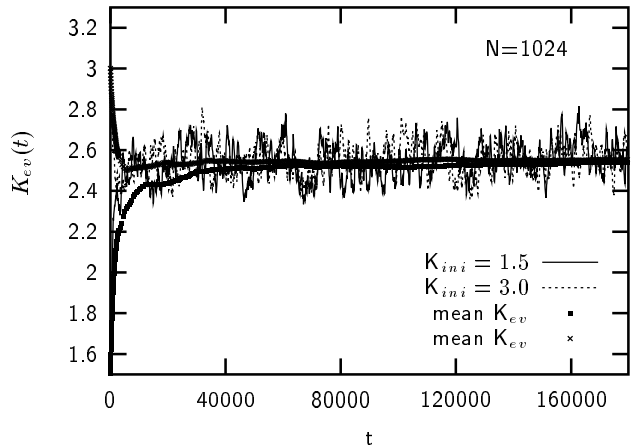}
	\caption{\small Evolution of the average connectivity $K_{ev}$ with an activity-driven rewiring, shown for two different initial connectivities. Independent of the initial conditions chosen at random, the networks evolve to an average connectivity $K_{ev}=2.55 \pm 0.04$. Time $t$ is in simulation steps.}
	\label{fig:rohlf-1-evolution}
\end{figure}

\begin{figure}[h!tbp]
\centering
\includegraphics[width=85mm]{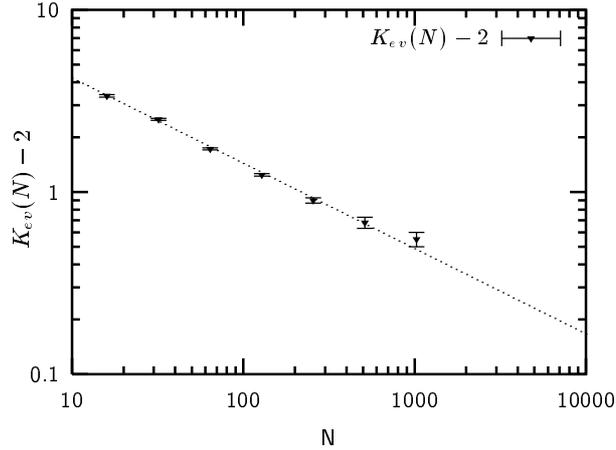}
\caption{\small A finite size scaling of the evolved average connectivity $K_{ev}$ as a function of network size $N$ reveals that for large $N$, the mean $K_{ev}$ converge towards the critical connectivity $K_c=2$. For systems with $N \leq 256$, the average was taken over $4 \times 10^6$ time steps, for $N = 512$ and $N = 1024$ over $5 \times 10^5$ and $2.5 \times 10^5$ time steps respectively.}
\label{fig:rohlf-2-finitesizeKev}
\end{figure}

A typical run of this algorithm will result in a connectivity evolution as shown in Figure~\ref{fig:rohlf-1-evolution} for a network of $N=1024$ nodes. Independent from the initial connectivity $K_{ini}$, the system evolves towards a statistically stationary state with an average evolved connectivity of $K_{ev}(N=1024) = 2.55 \pm 0.04$. With increasing system size $N$, $K_{ev}$ converges towards $K_c = 2$ for the large system limit $N \rightarrow \infty$ with a scaling relationship
\begin{equation}
	K_{ev} (N) = 2 + c N^{-\delta}
\end{equation}
with $c= 12.4 \pm 0.5$ and $\delta = 0.47 \pm 0.01$ (compare Figure~\ref{fig:rohlf-2-finitesizeKev}).

\begin{figure}[h!tbp]
	\centering
	\includegraphics[width=85mm]{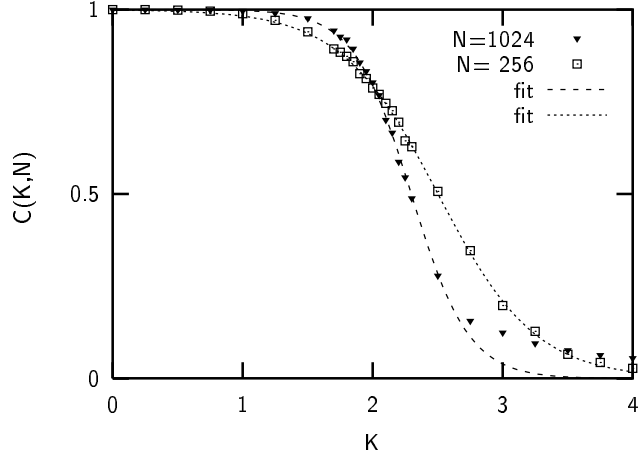}
	\caption{\small The frozen component $C(K,N)$ of random threshold
networks, as a function of the networks' average connectivities $K$, shown for two different network sizes $N$. Sigmoid function fits can be used for a finite size scaling of the transition connectivity from active to frozen networks. The results indicate that this transition takes place at $K_c=2$ for large $N$, details can be found in the original work \citep{Bornholdt:2000a}.}
	\label{fig:rohlf-3-frozencomponent}
\end{figure}

The underlying principle which facilitates self-organization in this model is based on the activity $A(i)$ of a node being closely connected to the frozen component of the network -- the fraction of nodes which do not change their state along the attractor -- which also undergoes a transition from a large to a vanishing frozen component at the critical connectivity. At low network connectivity, a large fraction of nodes will likely be frozen, and thus receive new in-links once they are selected for rewiring. On the other hand, at high connectivity, nodes will often change their state and in turn lose in-links in the rewiring process. Figure~\ref{fig:rohlf-3-frozencomponent} shows the above mentioned transition as a function of connectivity for two different network sizes. With a finite size scaling of the transition connectivities at the respective network sizes, it can be shown that for large $N \rightarrow \infty$, the transition occurs at the critical value of $K_c=2$.

\subsection{Correlation as a criterion for rewiring: self-organization on a spin lattice neural network model}\label{sec:BornholdtRoehl}

The models described in the following section and originally proposed
by \citet{Bornholdt:2001,Bornholdt:2003} capture self-organized
critical behavior on a two-dimensional spin lattice via a simple
correlation-based rewiring method. The motivation behind the new
approach was to transfer the idea of self-organization
from \citet{Bornholdt:2000a} to neural networks, thus creating a
first self-organized critical neural network model. 

In contrast to the activity-regulated model \citep{Bornholdt:2000a} discussed above, now:
\begin{itemize}
	\item the topology is constrained to a squared lattice,
	\item thermal noise is added to the system,
	\item link weights take non-discrete values, and
	\item activation thresholds may vary from 0.
\end{itemize}
The model consists of a randomly and asymetrically connected threshold network of $N$ spins ($\sigma_i = \pm 1$), where links can only be established among the eight local neighbors of any lattice site. The link weights $w_{ij}$ can be activating or inhibiting and are chosen randomly from a uniform distribution $w_{ij} \in \{-1,+1\}$. The average connectivity $K$ denotes the average number of non-zero incoming weights. All nodes are updated synchronously via a simple threshold function of their input signals from the previous time step:
\begin{eqnarray}
	\mathrm{Prob} [ \sigma_i(t+1) = +1 ] &=& g_\beta(f_i(t)) \nonumber \\
	\mathrm{Prob} [ \sigma_i(t+1) = -1 ] &=& 1- g_\beta(f_i(t))
\end{eqnarray} 
with
\begin{equation}
	f_i(t) = \sum_{j=1}^N w_{ij} \sigma_j(t) + \Theta_i
\end{equation}
and
\begin{equation}
	g_\beta(f_i(t))=\frac{1}{1+\exp(-2\beta f_i(t))}
\end{equation}
where $\beta$ denotes the inverse temperature and $\Theta_i$ is the activation threshold of node $i$. Thresholds are chosen as $\Theta_i = -0.1 + \gamma$ where $\gamma$ is a small random Gaussian noise of width $\epsilon$. The model per se exhibits a percolation transition under variation of $K$ or $\Theta$, changing between a phase of ordered dynamics, with short transients and limit cycle attractors, and a chaotic phase with cycle lengths scaling exponentially with system size.

On a larger time scale, the network topology is now changed by a slow local rewiring mechanism according to the following principle: if the dynamics of two neighboring nodes is highly correlated or anti-correlated, they get a new link between them. Otherwise, if their activity shows low correlation, any present link between them is removed, which is reminiscient of the Hebbian learning rule. In this model, the correlation $C_{ij}(\tau)$ of nodes $i,j$ over a time interval $\tau$ is defined as
\begin{equation}
	C_{ij}(\tau) = \frac{1}{\tau+1} \sum_{t=t_0}^{t_0+\tau} \sigma_i(t) \sigma_j(t).
\end{equation} 

The full model is constructed as follows:
\begin{enumerate}
\item Start with a randomly connected lattice of average connectivity $K_{ini}$, random initial node configuration, and random individual activation thresholds $\Theta_i$.
\item Perform synchronous updates of all nodes for $\tau$ time steps (Here, the choice of $\tau$ is not linked to any attractor period measurement, but should be chosen large enough to ensure a separation of time scales between network dynamics and topology changes.).
\item Choose a random node $i$ and random neighbor $j$ and calculate $C_{ij}(\tau/2)$ over the last $\tau/2$ time steps  (a first $\tau /2$ time steps are disregarded to exclude a possible transient period following the previous topology change.).
\item If $\vert C_{ij}(\tau/2) \vert$ is larger than a given threshold $\alpha$, a new link from node $j$ to $i$ is inserted with random weight $w_{ij} \in \{-1,+1\}$. If else $\vert C_{ij}(\tau/2) \vert \leq \alpha$ the weight $w_{ij}$ is set to 0.
	\item Use the current network configuration as new initial state and continue with step 2.
\end{enumerate}

\begin{figure}[h!tbp]
	\centering
	\includegraphics[width=0.505\textwidth]{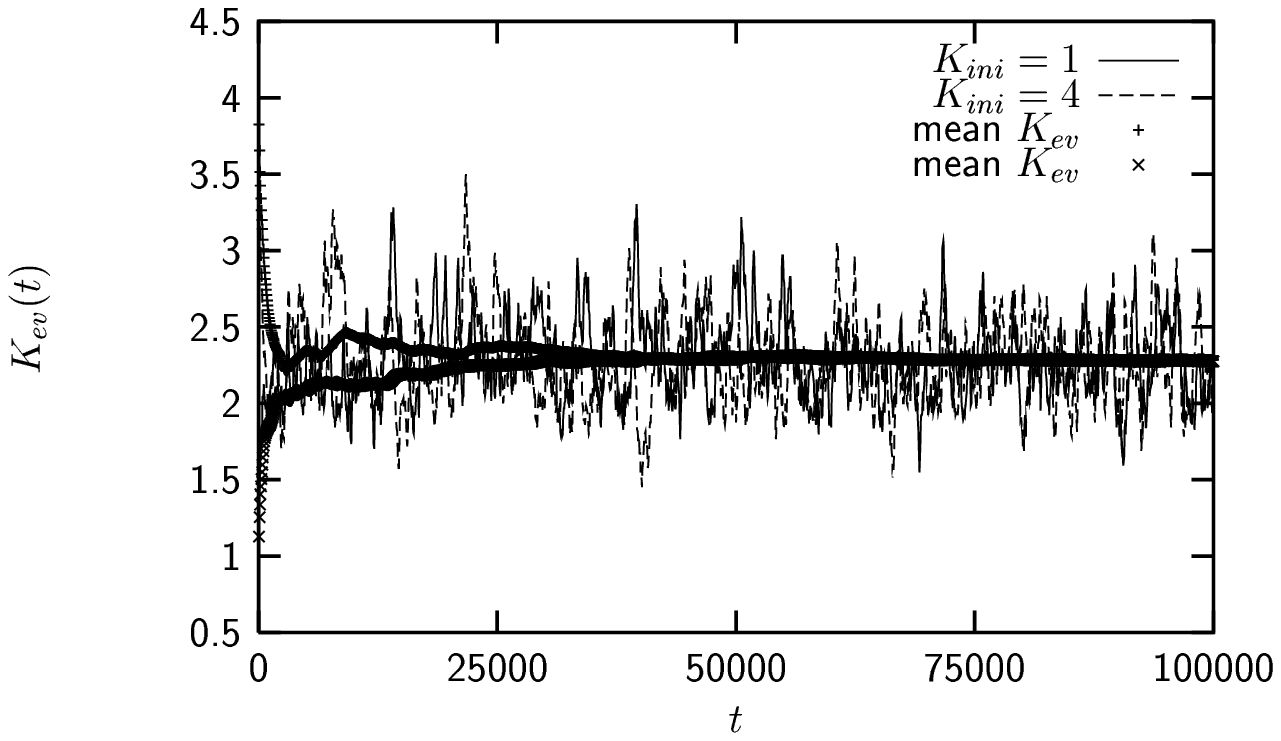}
	\includegraphics[width=0.48\textwidth]{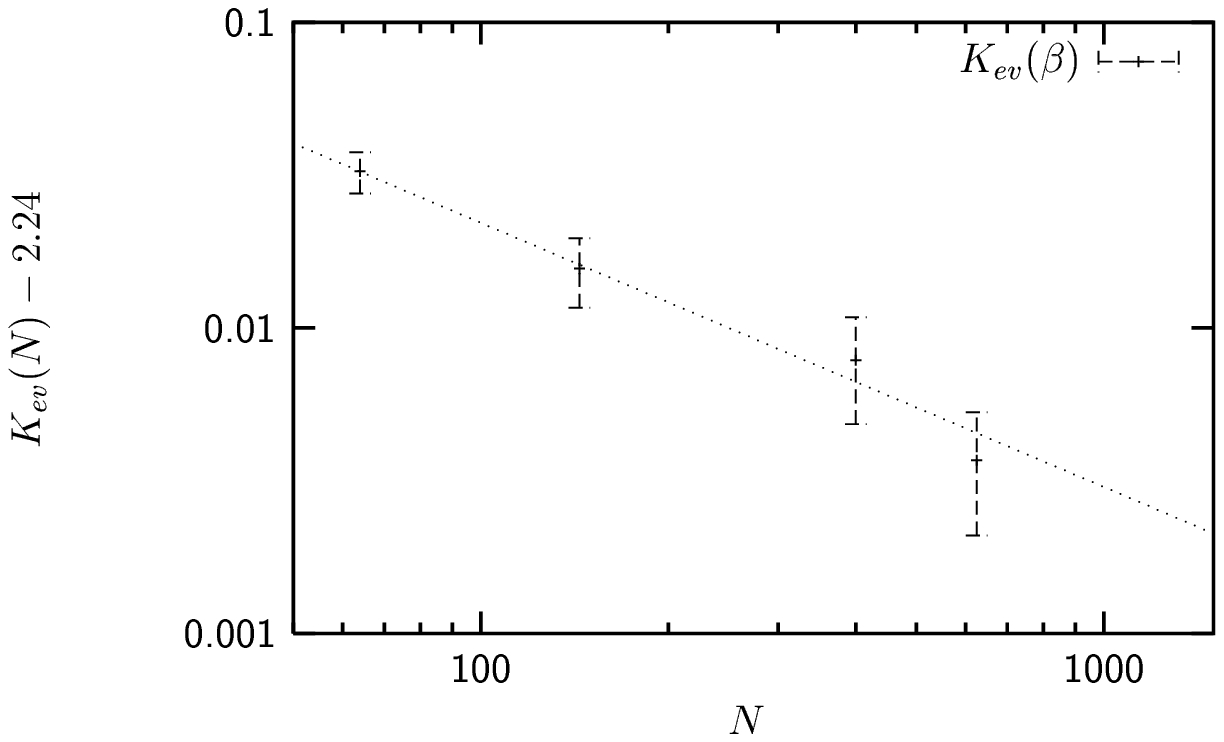}
	\caption{\small Left: Evolution of the average connectivity with a correlation-based rewiring, shown for two different initial connectivities. Again, connectivity evolves to a specific average depending on network size, but independent from the initial configuration. Right: Finite size scaling of the evolved
average connectivity. The best fit is obtained for a large system limit of $K_{ev}^\infty = 2.24 \pm 0.03$.
Averages are taken over $4 \times 10^5$ time steps.}
	\label{fig:roehl-1-evolution-2-finitesizeKev}
\end{figure}

Independent from the initial average connectivity $K_{ini}$, one finds a slow convergence of $K$ towards a specific mean evolved connectivity $K_{ev}$, which is characteristic for the respective network size $N$ (Figure~\ref{fig:roehl-1-evolution-2-finitesizeKev}) and shows a finite size scaling according to
\begin{equation}
	K_{ev}(N) = aN^{-\delta}+b
\end{equation}
with $a = 1.2 \pm 0.4$, $\delta = 0.86 \pm 0.07$ and $b = 2.24 \pm 0.03$, where $b$ can be interpreted as the evolved average connectivity for the large system limit $N \rightarrow \infty$:
\begin{equation}
	K_{ev}^\infty = 2.24 \pm 0.03
\end{equation}

\begin{figure}[h!tbp]
	\centering
	\includegraphics[width=0.49\textwidth]{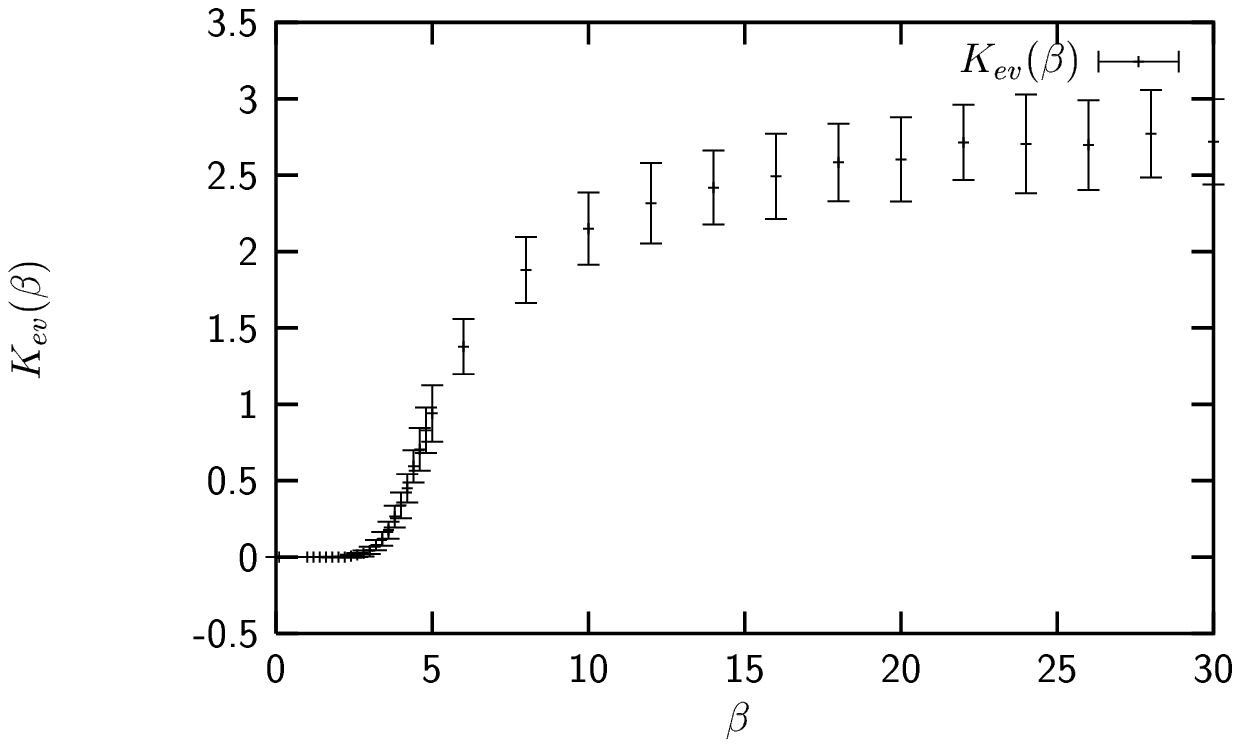}
	\includegraphics[width=0.49\textwidth]{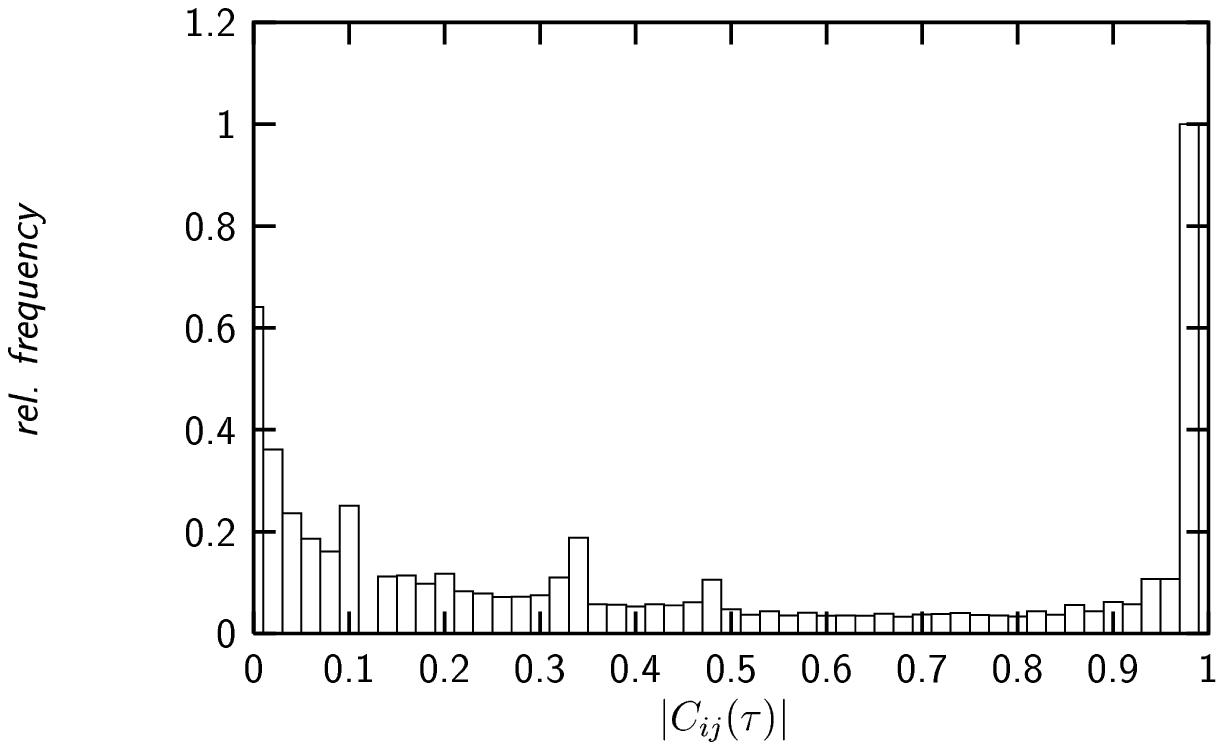}
	\caption{\small Left: Evolved average connectivity $K_{ev}$ as a function of
the inverse temperature $\beta$. The behavior is robust over a wide range of $\beta$. Each point is averaged over $10^5$ time steps in a network of size $N = 64$. 
Right: Histogram of the average correlation $\mid C_{ij}(\tau)\mid $ for a network evolving in time with $N=64$ and $\beta=10$. As very low and very high correlations dominate, the exact choice of the correlation threshold during the rewiring process is of minor importance.}
	\label{fig:roehl-3-betaKev-5-alphahistogram}
\end{figure}

In addition, it is shown that the proposed adaptation mechanism works robustly towards a wide range of thermal noise $\beta$, and also the specific choice of the correlation threshold $\alpha$ for rewiring only plays a minor role regarding the evolved $K_{ev}$ (Figure~\ref{fig:roehl-3-betaKev-5-alphahistogram}).

\begin{figure}[h!tbp]
	\centering
	\includegraphics[width=85mm]{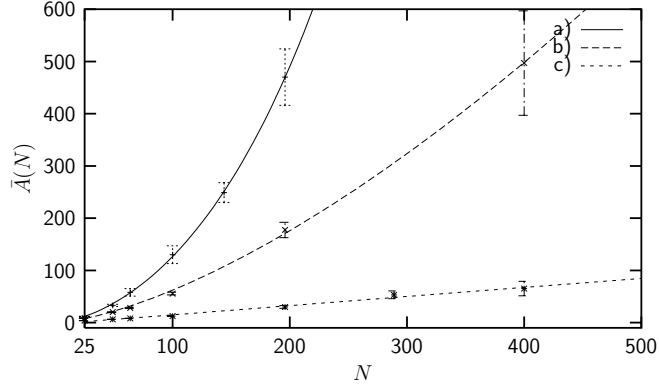}
	\caption{\small Finite size scaling of the evolved average attractor period $\bar{A}(N)$ for networks of different sizes $N$. 
(b). Also shown for comparison is the corresponding scaling of the
attractor lengths of a supercritical random network (a) with $K=3.8$
and a subcritical one (c) with $K=1.5$.
}
	\label{fig:roehl-7-attractorscaling}
\end{figure}

Having a closer look at a finite size scaling of the evolved average attractor length, one finds a scaling behavior close to criticality. While attractor lengths normally scale exponentially with system size in the supercritical, chaotic regime and sublinearly in the subcritical, ordered phase, this model exhibits relatively short attractor cycles also for large evolved networks in the critical regime (Figure~\ref{fig:roehl-7-attractorscaling}).

\subsection{Simplicity vs. biological plausibility -- and possible improvements}\label{sec:RybarschBornholdt01}

\subsubsection{Transition from spins to Boolean node states}

In the above sections, we have seen that already basic toy models, neglecting a lot of details, can reproduce some of the observations made in real neuronal systems. We now want to move these models a step closer towards biologigal plausibility and at the same time construct an even simpler model.

One major shortcoming of both models discussed above is the fact that they are constructed as spin models. In some circumstances, however, when faithful representation of certain biological details is important, the exact definition matters. In the spin version of a neural network model, for example, a node with negative spin state $\sigma_j=-1$ will transmit non-zero signals through its outgoing weights $c_{ij}$, despite representing an inactive (!) biological node. In the model, such signals arrive at target nodes $i$, e.g., as a sum of incoming signals $f_i= \sum_{j=1}^{N}c_{ij}\sigma_j$. However, biological nodes, as genes or neurons, usually do not transmit signals when inactive. Also in other contexts like biochemical network models, each node represents whether a specific chemical component is present $(\sigma=1)$ or absent $(\sigma=0)$. Thus the network itself is mostly in a state of being partially absent as, e.g., in a protein network where for every absent protein all of its outgoing links are absent as well. In the spin state convention, this fact is not faithfully represented. A far more natural choice would be to construct a model based on Boolean state nodes, where nodes can be truly "off" ($\sigma_i=0$) or "on" ($\sigma_i=1$) -- which is precisely what we are going to do in the following sections.

Another example for an inaccurate detail is the common practice to use the standard convention of the Heaviside step function as an activation function in discrete dynamical networks (or the sign function in the spin model context). The convention $\Theta(0)=1$ is not a careful representation of biological circumstances. Both, for genes and neurons, a silent input frequently maps to a silent output. Therefore, we use a redefined threshold function defined as 
\begin{equation} 
\Theta_0(x)= 
\left\{ 
\begin{array}{c} 
1, \ \ \  x>0 \noindent \\  
0, \ \ \ x \leq 0.  
\end{array} 
\right. 
\end{equation}  

Most importantly, in our context here, 
the choice of Boolean node states and the redefined threshold function are vital when we discuss mechanisms of self-organization. For instance, the correlation-based approach presented in the older spin model \citep{Bornholdt:2003} in section \ref{sec:BornholdtRoehl} explicitly measures contributions by pairs of nodes which could be constantly off ($\sigma_{i/j} = -1$) and still treat them as highly correlated (because $(-1) \cdot (-1) = +1$) even though there is no activity involved at all. In the later section \ref{sec:RybarschBornholdtEvolution}, we will therefore present a new approach for a network of Boolean state nodes, which does not rely on non-activity correlations anymore.

\subsubsection{Model definitions}
Let us consider randomly wired threshold networks of $N$ nodes $\sigma_i \in \{ 0, 1 \}$. At each discrete time step, all nodes are updated in parallel according to
\begin{equation}
	\sigma_i(t+1) = \Theta_0(f_i(t)) 
	\label{eq:SignumFunction}
\end{equation}
using the input function
\begin{equation}
	f_i(t) = \sum_{j=1}^N c_{ij} \sigma_j(t) + \theta_i.
	\label{eq:InputSum}
\end{equation}
In particular we choose $\Theta_0(0) := 0$ for plausibility reasons (zero input signal will produce zero output). While the weights take discrete values $c_{ij} = \pm 1$ with equal probability for connected nodes, we select the thresholds $\theta_i = 0$ for the following discussion. For any node $i$, the number of incoming links $c_{ij} \neq 0$ is called the in-degree $k_i$ of that specific node. ${K}$ denotes the average connectivity of the whole network. With randomly placed links, the probability for each node to actually have $k_i = k$ incoming links follows a Poissonian distribution:
\begin{equation}
	p(k_i=k) = \frac{{K}^k}{k!} \cdot \exp(-{K}).
	\label{eq:PoissonianLinkDistribution}
\end{equation}

\subsubsection{Exploration of critical properties -- Activity-dependent criticality}
To analytically derive the critical connectivity of this type of network model, we first study damage spreading on a local basis and calculate the probability $p_s(k)$ for a single node to propagate a small perturbation, i.e. to change its output from 0 to 1 or vice versa after changing a single input state. The calculation can be done closely following the derivation for spin-type threshold networks by \citet{Rohlf:2002}, but one has to account for the possible occurrence of `0' input signals also via non-zero links. The combinatorial approach yields a result that directly corresponds to the spin-type network calculation via $p_s^{\mathrm{bool}}(k) = p_s^{\mathrm{spin}}(2k)$.
\begin{figure}[h!tbp]
	\centering
	\includegraphics[width=85mm]{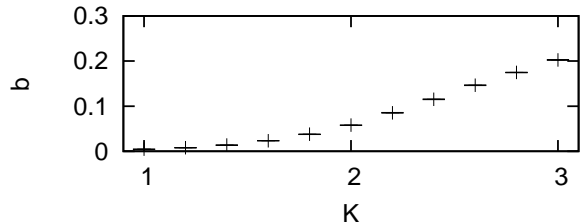}
	\caption{\small Average node activity $b$ as function of connectivity $K$ measured on attractors of 10000 sample networks each, 200 nodes.}
	\label{fig:rybarsch-1-activity}
\end{figure}
However, this approach does not hold true for our Boolean model in combination with the defined Theta function $\Theta_0(0) := 0$ as it assumes a statistically equal distribution of all possible input configurations for a single node. In the Boolean model, this would involve an average node activity of $b=0.5$ over the whole network (where $b$ denotes the average fraction of nodes which are active, i.e. $\sigma_i=1$). Instead we find (Fig.~\ref{fig:rybarsch-1-activity}) that the average activity on the network is significantly below $0.5$. At $K=4$ (which will turn out to be already far in the supercritical regime), less than 30 percent of all nodes are active on average. Around $K\approx 2$ (where we usually expect the critical connectivity for such networks), the average activity is in fact below 10 percent. Thus, random input configurations of nodes in this network will more likely consist of a higher number of `0' signal contributions than of $\pm 1$ inputs.

Therefore, when counting input configurations for the combinatorial
derivation of $p_s(k)$, we need to weight all relevant configurations
according to their realization probability as given by the average
activity $b$ -- the detailed derivation can be found
in \citep{Rybarsch:2012a}. With the average probability of damage
spreading, we can further compute the branching parameter or
sensitivity $\lambda = K \cdot \langle p_s \rangle ({K})$ and apply
the annealed approximation \citep{Derrida:1986, Bornholdt:2000a} to
obtain the critical connectivity $K_c$ by solving 
\begin{equation}
	\lambda_c = K_c \cdot \langle p_s \rangle (K_c)  = 1.
	\label{eq:K_cCalculation}
\end{equation}

However, $K_c$ now depends on the average network activity, which in turn is a function of the average connectivity ${K}$ itself as shown in Fig.~\ref{fig:rybarsch-1-activity}.
\begin{figure}[h!tbp]
	\centering
	\includegraphics[width=85mm]{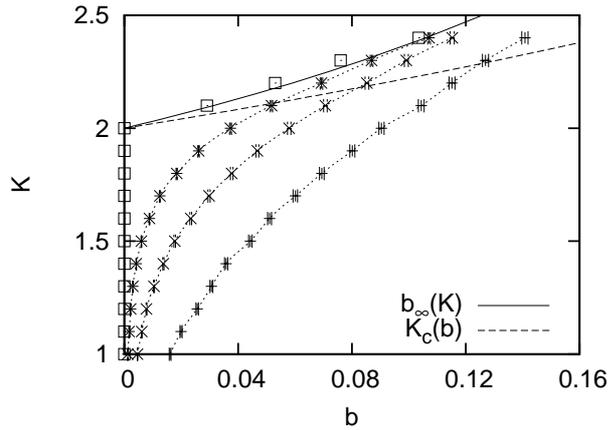}
	\caption{\small Average activity $b$ on attractors of different network sizes (right to left: $N=50, 200, 800$, ensemble averages were taken over 10000 networks each). Squares indicate activity on an infinite system determined by finite size scaling, which is in good agreement with the analytic result (solid line). The dashed line shows the analytic result for $K_c(b)$ from eq.\ (\ref{eq:K_cCalculation}). The intersections represent the value of $K_c$ for the given network size.}
	\label{fig:rybarsch-2-activityVsKc}
\end{figure}
From the combined plot in Fig.~\ref{fig:rybarsch-2-activityVsKc} we find that both curves intersect at a point where the network dynamics -- due to the current connectivity $K$ -- exhibit an average activity which in turn yields a critical connectivity $K_c$ that exactly matches the given connectivity. This intersection thus corresponds to the critical connectivity of the present network model. 

However, the average activity still varies with different network sizes, which is obvious from Figure~\ref{fig:rybarsch-2-activityVsKc}. Therefore, also the critical connectivity is a function of $N$. 
For an analytic approach to the infinite size limit, it is possible to calculate the average network activity at the next time step $b_{t+1}$ in an iterative way from the momentary average activity $b_t$. Again, the details can be found in \citep{Rybarsch:2012a}. We can afterwards distinguish between the different dynamical regimes by solving $\langle b_{t+1} \rangle = b_t(K)$ for the critical line. The solid line in Figure~\ref{fig:rybarsch-2-activityVsKc} depicts the evolved activity in the long time limit. We find that for infinite system size, the critical connectivity is at 
\begin{equation*}
{K}_c(N \rightarrow \infty) = 2.000 \pm 0.001
\end{equation*}
while up to this value all network activity vanishes in the long time limit ($b_\infty=0$). For any average connectivity ${K} > 2$, a certain fraction of nodes remains active. In finite size systems, both network activity evolution and damage propagation probabilities are subject to finite size effects, thus increasing $K_c$ to a higher value.

\begin{figure}[h!tbp]
	\centering
	\includegraphics[width=85mm]{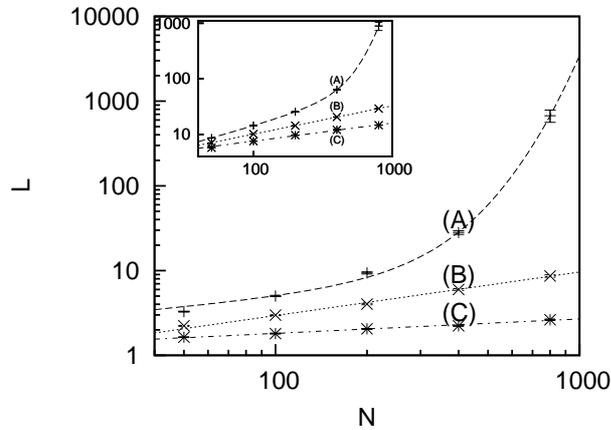}
	\caption{\small Average attractor length at different network sizes. Ensemble averages were taken over 10000 networks each at (A) $K=2.4$, (B) $K=2.0$, (C) $K=1.6$. Inset figure shows the scaling behavior of the corresponding transient lengths.}
	\label{fig:rybarsch-4-attractors}
\end{figure}
Finally, let us have a closer look on the average length of attractor cycles and transients. As shown in Fig.~\ref{fig:rybarsch-4-attractors}, the behavior is strongly dependent of the dynamical regime of the network. As expected and in accordance with early works on random threshold networks \citep{Kurten:1988} as well as random Boolean networks \citep{Bastolla:1996}, we find an exponential increase of the average attractor lengths with network size $N$ in the chaotic regime ($K>K_c$), whereas we can observe a power-law increase in the frozen phase ($K<K_c$). We find similar behavior for the scaling of transient lengths (inset of Figure~\ref{fig:rybarsch-4-attractors}).

\subsubsection{Extension of the model: thermal noise}\label{sec:RybarschBornholdt01-noise}

As it is clear from eq.~\eqref{eq:SignumFunction}, nodes in our model
will only switch to active state if they get a positive input
from \emph{somewhere}. Thus, to get activity into the system, we could
either define certain nodes to get an additional external input, but
this would at the same time create two different kinds of nodes, those
with and those without external activity input, which would in turn
diminish the simplicity of our model. That is why we will
alternatively use thermal noise to create activity, using a Glauber
update of the nodes in the same way as it was discussed in the spin
model \citep{Bornholdt:2003} in section \ref{sec:BornholdtRoehl},
with one slight modification. We define a probability for the
activation of a node, which is a sigmoid function of the actual input
sum, but also leaving room for a spontaneous activation related to the
inverse temperature $\beta$ of the system. 
\begin{eqnarray}\label{eq:01-update}
	\mathrm{Prob} [ \sigma_i(t+1) = 1 ] &=& g_\beta(f_i(t)) \nonumber \\
	\mathrm{Prob} [ \sigma_i(t+1) = 0 ] &=& 1- g_\beta(f_i(t))
\end{eqnarray} 
with
\begin{equation}
	f_i(t) = \sum_{j=1}^N c_{ij} \sigma_j(t) + \Theta_i 
\end{equation}
and
\begin{equation}\label{eq:01-activation-probability-input-shift}
	g_\beta(f_i(t))=\frac{1}{1+\exp(-2\beta (f_i(t) -0.5 ))}.
\end{equation}
\begin{figure}[h!tbp]
	\centering
	\includegraphics[width=85mm]{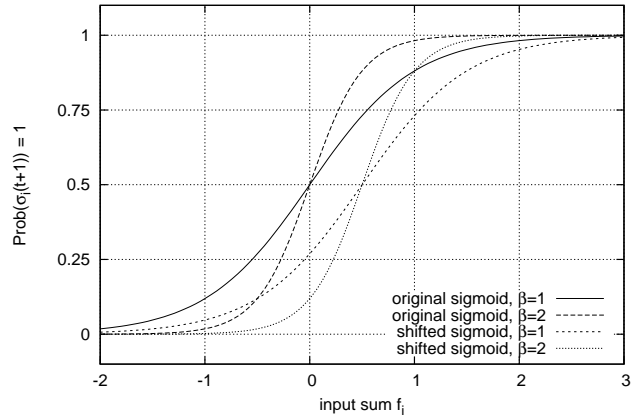}
	\caption{\small With integer input sum values, nodes without input would turn active with probability $0.5$ regardless of temperature. To prevent this, we shift the input sum by $-0.5$ such that the transition between off and on state happens exactly in the middle between input sums 0 resp. 1.}
	\label{fig:01-shiftedsigmoid}
\end{figure}
You will note the similarity to the older spin
model \citep{Bornholdt:2003}, but be aware that in
eq.~\eqref{eq:01-activation-probability-input-shift} we shift the
input sum $f_i$ by $-0.5$, and we will explain now why this is
necessary. Remember that we use binary coupling weights $c_{ij}=\pm 1$
for existing links in our model. The input sum $f_i$ to any node will
therefore be an integer value. If we would not shift the input sum, a
node with an input $f_i=0$ (which should always be inactive in the
deterministic model without thermal noise), would, after the
introduction of the Glauber update rule with non-zero temperature,
always have a probability of $\mathrm{Prob} [ \sigma_i(t+1) = +1 ] =
0.5$ to be activated, regardless of the actual inverse temperature
$\beta$. Figure~\ref{fig:01-shiftedsigmoid} illustrates this
problem. A simple shift of $-0.5$ will now give us the desired
behavior: Nodes with input $f_i=0$ will stay off in most cases, with a
slight probability to turn active depending on temperature, and on the
other hand nodes with activating input $f_i=+1$ will be on in most
cases, with slight probability to remain inactive. With this
modification of the original basic model \citep{Rybarsch:2012a}, we can
now continue to make our network adaptive and critical.

\subsection{Self-organization on the Boolean state model}\label{sec:RybarschBornholdtEvolution}

We now want to set up an adaptive network based on the model discussed above, which is still capable of self-regulation towards a critical state despite being simplified to the most minimal model possible. Particularly, we want it to
\begin{itemize}
	\item have a simple, yet biologically plausible rewiring rule, which only uses local information accessible to individual nodes
	\item work independently from a special topology as a lattice.
\end{itemize}

We have already pointed out the problems of a spin formulation of neural network models like \citet{Bornholdt:2003}, and possible ways out with the Boolean state model. As a major advantage of the latter, activity avalanches intrinsically occur in this type of network, whereas spin networks typically exhibit continuous fluctuations with no avalanches directly visible. However, the old correlation-based rewiring mechanism will no longer work when inactive nodes are now represented by `$0$' instead of `$-1$'. A solution will be presented below.

A second aspect which needs to be addressed concerning the self-organization mechanism is topological invariance of the algorithm. The older, correlation-based mechanism from the spin model relies on randomly selecting neighboring sites on a lattice for the rewiring processes. On a lattice, the number of possible neighbors is strictly limited, but on a large random network near critical connectivity, there are far more unconnected pairs of nodes than there are connected pairs. Thus, randomly selecting pairs of nodes for rewiring would introduce a strong bias towards connecting nodes which were previously unconnected. This results in a strong increase of connectivity and makes a self-organized adaptation towards a critical state practically impossible.
If we want to overcome the restriction of e.g.\ a confined lattice topology in order to improve biological applicability of the model, we have to adapt the rewiring mechanism such that this bias no longer exists.

Of course, the new model shall inherit several important features from the older spin models, which already underline the applicability to biological networks: in particular, it must be capable of self-regulation towards a critical state despite being simplified to the most minimal model possible. The organization process however should be based on a simple, yet biologically plausible rewiring rule, which only uses local information accessible to individual nodes like pre-\ and post-synaptic activity and correlation of such activity.

This section is directly based on our recent work \citep{Rybarsch:2012b}.

\subsubsection{Model definitions}

Consider a randomly connected threshold network of the type discussed above in section \ref{sec:RybarschBornholdt01}. The network consists of $N$ nodes of Boolean states $\sigma_i \in \{ 0, 1 \}$ which can be linked by asymmetric directed couplings $c_{ij} = \pm 1$. Node pairs which are not linked have their coupling set to $c_{ij} = 0$; and links may exist between any two nodes, so there is no underlying spatial topology in this model.

All nodes are updated synchronously in discrete time steps via a simple threshold function of their input signals with a little thermal noise introduced by the inverse temperature $\beta$, in the same way as in the model of Bornholdt and Roehl (section \ref{sec:BornholdtRoehl}), but now with an input shift of $-0.5$ in the Glauber update, representing the new $\Theta_0$ function as discussed in section \ref{sec:RybarschBornholdt01-noise}:
\begin{eqnarray}
	\mathrm{Prob} [ \sigma_i(t+1) = 1 ] &=& g_\beta(f_i(t)) \nonumber \\
	\mathrm{Prob} [ \sigma_i(t+1) = 0 ] &=& 1- g_\beta(f_i(t))
\end{eqnarray} 
with
\begin{equation}
	f_i(t) = \sum_{j=1}^N c_{ij} \sigma_j(t) + \Theta_i 
\end{equation}
and
\begin{equation}
	g_\beta(f_i(t))=\frac{1}{1+\exp(-2\beta (f_i(t)-0.5))}.
\end{equation}
For the simplicity of our model, we first assume that all nodes have an identical activation threshold of $\Theta_i = 0$.

\subsubsection{Rewiring algorithm}

The adaptation algorithm is now constructed in the following way: We start the network at an arbitrary initial connectivity $K_{ini}$ and do parallel synchronous updates on all nodes according to eq. \eqref{eq:01-update}. All activity in this model originates from small perturbations by thermal noise, leading to activity avalanches of various sizes. In this case we set the inverse temperature to $\beta=5$. On a larger time scale, i.e. after $\tau=200$ updates, a topology rewiring is introduced at a randomly chosen, single node. The new element in our approach is to test whether the addition or the removal of one random in-link at the selected node will increase the average dynamical correlation to all existing inputs of that node. By selecting only one single node for this procedure, we effectively diminish the bias of selecting mostly unconnected node pairs -- but retain the biologically inspired idea for a Hebbian, correlation-based rewiring mechanism on a local level.

Now, we have to define what is meant by \emph{dynamical correlation} in this case. We here use the Pearson correlation coefficient to first determine the dynamical correlation between a node $i$ and one of its inputs $j$:
\begin{equation}\label{eq:PearsonCorrelation}
	C_{ij} = \frac{\langle \sigma_i(t+1) \sigma_j(t) \rangle - \langle \sigma_i(t+1) \rangle \langle \sigma_j(t) \rangle}{S_i \cdot S_j}
\end{equation}
where $S_{i}$ and $S_{j}$ in the denominator depict the standard deviation of states of the nodes $i$ and $j$, respectively. In case one or both of the nodes are frozen in their state (i.e. yield a standard deviation of 0), the Pearson correlation coefficient would not be defined, we will assume a correlation of $C_{ij}=0$. Note that we always use the state of node $i$ at one time step later than node $j$, thereby taking into account the signal transmission time of one time step from one node to the next one. Again, as in the model of \citet{Bornholdt:2003}, the correlation coefficient is only taken over the second half of the preceding $\tau$ time steps in order to avoid transient dynamics.
Finally, we define the average input correlation $C_i^{avg}$ of node $i$ as
\begin{equation}\label{eq:AverageInputCorrelation}
	C_i^{avg} = \frac{1}{k_i} \sum_{j=0}^N \vert c_{ij} \vert C_{ij}
\end{equation}
where $k_i$ is the in-degree of node $i$. The factor $\vert c_{ij} \vert$ ensures that correlations are only measured where links are present between the nodes. For nodes without any in-links ($k_i = 0$) we define $C_i^{avg} := 0$.

In detail, the adaptive rewiring is now performed in the following steps:
\begin{enumerate}
	\item Select a random node $i$ and generate three clones of the current network topology and state:
	\begin{description}
		\item[Network 1:] This copy remains unchanged.
		\item[Network 2:] In this copy, node $i$ will get one additional in-link from a randomly selected other node $j$ which is not yet connected to $i$ in the original copy (if possible, i.e. $k_i<N$). Also the coupling weight $c_{ij} \pm 1$ of the new link is chosen at random. 
		\item[Network 3:] In the third copy, one of the existing in-links to node $i$ (again if possible, i.e. $k_i>0$) will be removed.
	\end{description}
	\item All three copies of the network are individually run for $\tau=200$ time steps.
	\item On all three networks, the average input correlation $C_i^{avg}$ of node $i$ to all of the respective input nodes in each network is determined.
	\item We accept and continue with the network which yields the highest absolute value of $C_i^{avg}$, the other two clones are dismissed. If two or more networks yield the same highest average input correlation such that no explicit decision is possible, we simply continue with the unchanged status quo.
	\item A new random node $i$ is selected and the algorithm starts over with step 1.
\end{enumerate}

\begin{figure}[h!tbp]
\centering
\includegraphics[width=0.99\textwidth]{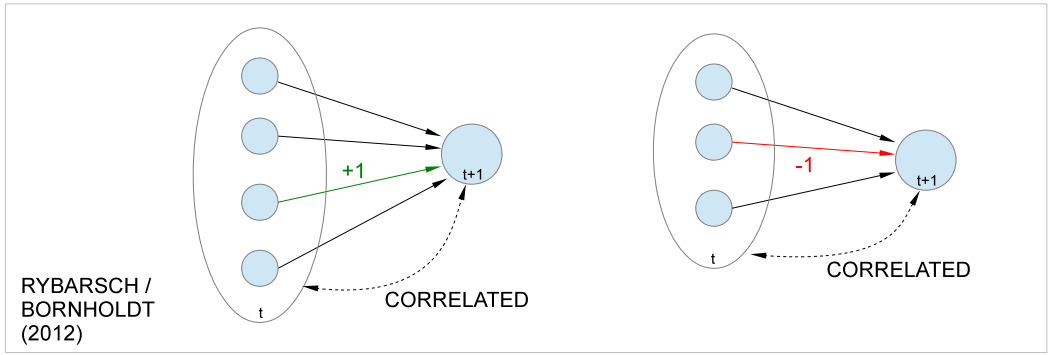}
\vspace*{-3ex}
\caption{\small Schematic illustration of the rewiring mechanism based on average input correlation. In this example, the target node initially has three in-links. Left: If the addition of a fourth input increases the average input correlation $C_{i}^{avg}$, a link will be inserted. Right: If removal of an existing in-link increases $C_{i}^{avg}$, the link will be deleted.}
\label{fig:01-evolution-schematic}
\end{figure}

It is worth noting that this model -- in the same way as the earlier work by \citet{Bornholdt:2003} -- uses locally available information at synapse level and takes into account both pre- and post-synaptic activity. This is a fundamental difference to approaches discussed e.g.\ by \citet{Arcangelis:2006}, \citet{Pellegrini:2007} or \citet{Levina:2007}, where only pre-synaptic activity determines changes to the coupling weights.

Note that the non-locality of running three network copies in parallel
that we use here is not critical. A local implementation is
straightforward, locally estimating the order parameter (here the
average input correlation $C_i^{avg}$) as time average, with a
sufficient time scale separation towards the adaptive changes in the
network. A local version of the model is studied in the latest version
of \citet{Rybarsch:2012b}.  

\subsubsection{Observations}
In the following, we will have a look at different observables during numerical simulations of network evolution in the model presented above. Key figures include the average connectivity $K$ and the branching parameter (or sensitivity) $\lambda$. Both are closely linked to and influenced by the ratio of activating links $p$, which is simply the fraction of positive couplings $c_{ij}=+1$ among all existing (non-zero) links.

\begin{figure}[h!tbp]
	\centering
	\includegraphics[width=0.49\textwidth]{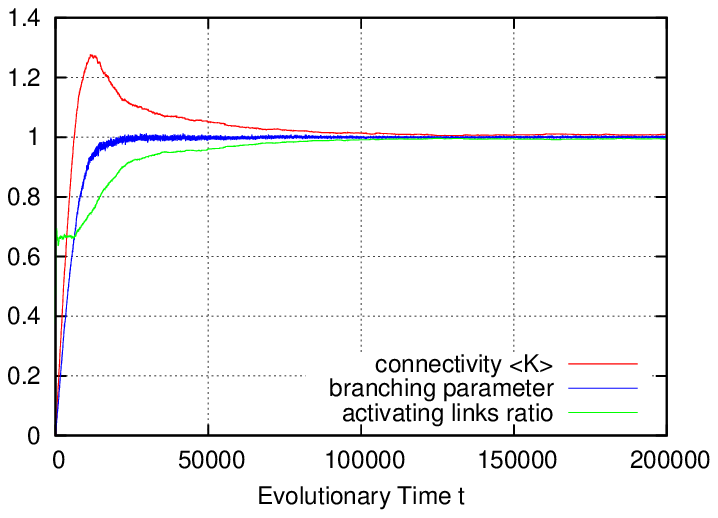}
	\includegraphics[width=0.49\textwidth]{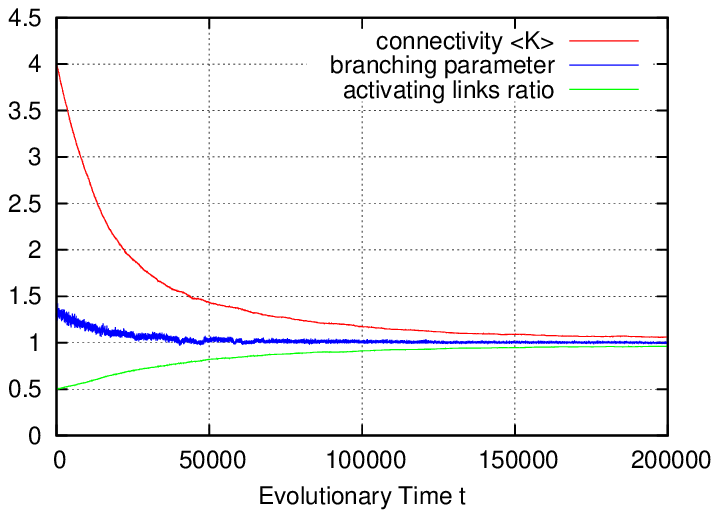}
	\caption{\small Regardless of initial connectivity and sensitivity, the network evolves to a critical configuration. Left: starting with completely disconnected nodes and obviously subcritical ``network''. Right: starting with a supercritical network.}
	\label{fig:01-evolution}
\end{figure}

The left part in Figure \ref{fig:01-evolution} shows an exemplary run of the topology evolution algorithm, where we start with completely isolated nodes without any links. Trivially, the ``network'' is subcritical at this stage, which can be seen from the branching parameter which is far below 1. As soon as rewiring takes place, the network starts to insert new links, obviously because these links enable the nodes to pass signals and subsequently act in a correlated way. With increasing connectivity, also the branching parameter rises, indicating that perturbations start to spread from their origin to other nodes. When the branching parameter approaches 1, indicating that the network reaches a critical state, the insertion of new links is cut back. The processes of insertion and depletion of links tend to be balanced against each other, regulating the network close to criticality.

If we, on the other hand, start with a randomly interconnected network at a higher connectivity like $K_{ini}= 4$ as shown in the right-hand side of Figure \ref{fig:01-evolution}, we find the network in the supercritical regime ($\lambda > 1$)
 at the beginning. When above the critical threshold, many nodes will show chaotic activity with very low average correlation to their respective inputs. The rewiring algorithm reacts in the appropriate way by deleting links from the network, until the branching parameter approaches 1.
 
In both examples above, the evolution of the ratio of activating links $p$ (which tends towards 1) shows, that the rewiring algorithm in general favors the insertion of activating links and vice versa the deletion of inhibitory couplings. This appears indeed very plausible if we remind ourselves that the rewiring mechanism optimizes the inputs of a node towards high correlation on average. Also, nodes will only switch to active state $\sigma_i=1$ if they get an overall positive input. As we had replaced spins by Boolean state nodes, this can only happen via activating links -- and that is why correlations mainly originate from positive couplings in our model. As a result, we observe the connectivity evolving towards one in-link per node, with the ratio of positive links also tending towards one.

\begin{figure}[h!tbp]
	\centering
	\includegraphics[width=0.49\textwidth]{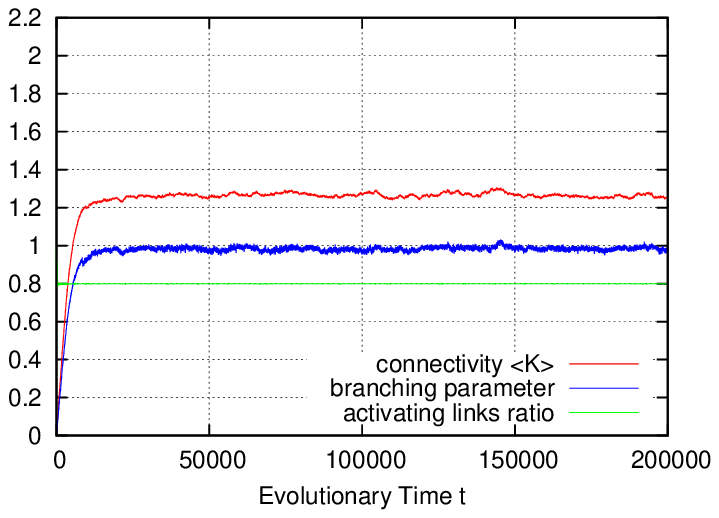}
	\includegraphics[width=0.49\textwidth]{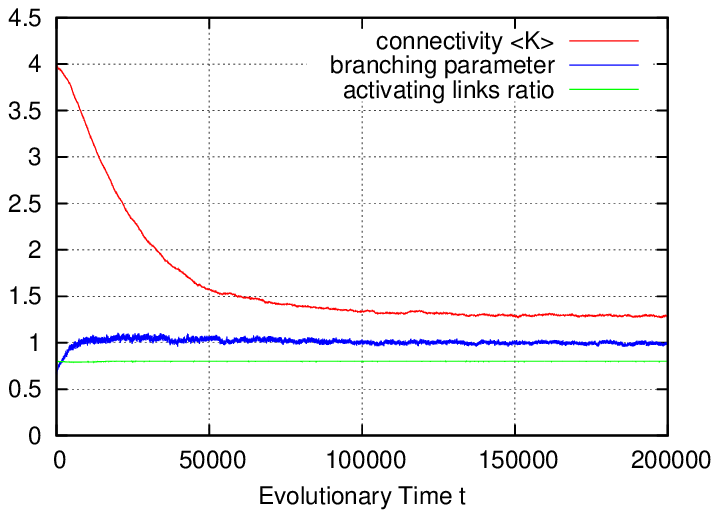}
	\caption{\small Network evolution with activating links ratio kept at $p=0.8$.}
	\label{fig:01-evolution-reshuffled}
\end{figure}

For a richer pattern complexity, we might later want to introduce a
second mechanism which balances out positive via negative links, and
with a first approach we can already test how the rewiring strategy
would fit to that situation: if, after each rewiring step, we change
the sign of single random links as necessary to obtain a ratio of
e.g.\ 80\% activating links (i.e.\ $p=0.8$), keeping the large
majority of present links unchanged, the branching parameter will
again stabilize close to the critical transition, while the
connectivity is maintained at a higher
value. Figure~\ref{fig:01-evolution-reshuffled} shows that the
self-organization behavior is again independent from the initial
conditions. This result does not depend on the exact choice of the
activating links ratio $p$; similar plots can easily be obtained for a
large range starting at $p = 0.5$, where the connectivity will
subsequently evolve towards a value slightly below $K=2$, which is the
value we would expect as the critical connectivity for a randomly
wired network with balanced link ratio according to the calculations
made in section \ref{sec:RybarschBornholdt01} for the basic network
model \citep{Rybarsch:2012a}. 

\begin{figure}[h!tbp]
	\centering
	\includegraphics[width=0.49\textwidth]{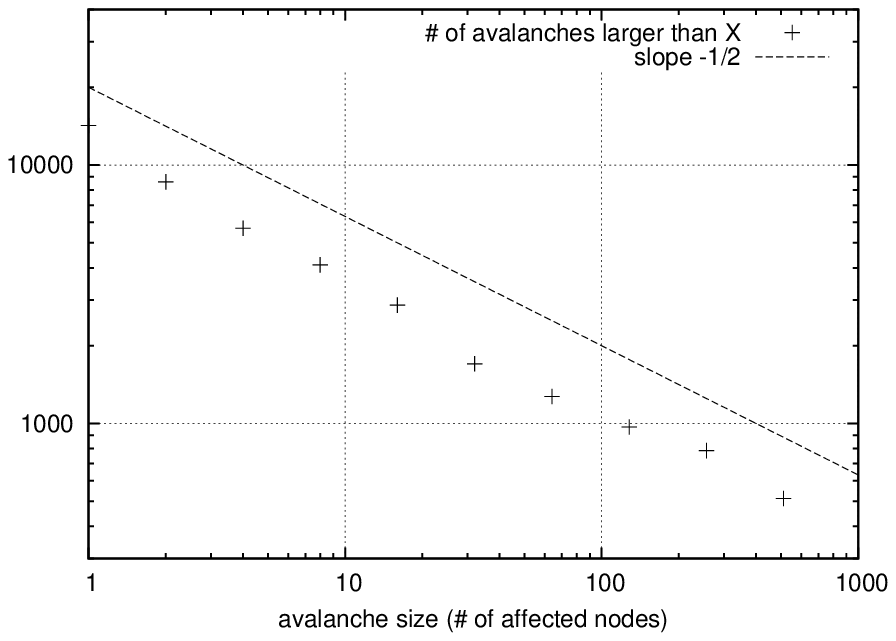}
	\includegraphics[width=0.49\textwidth]{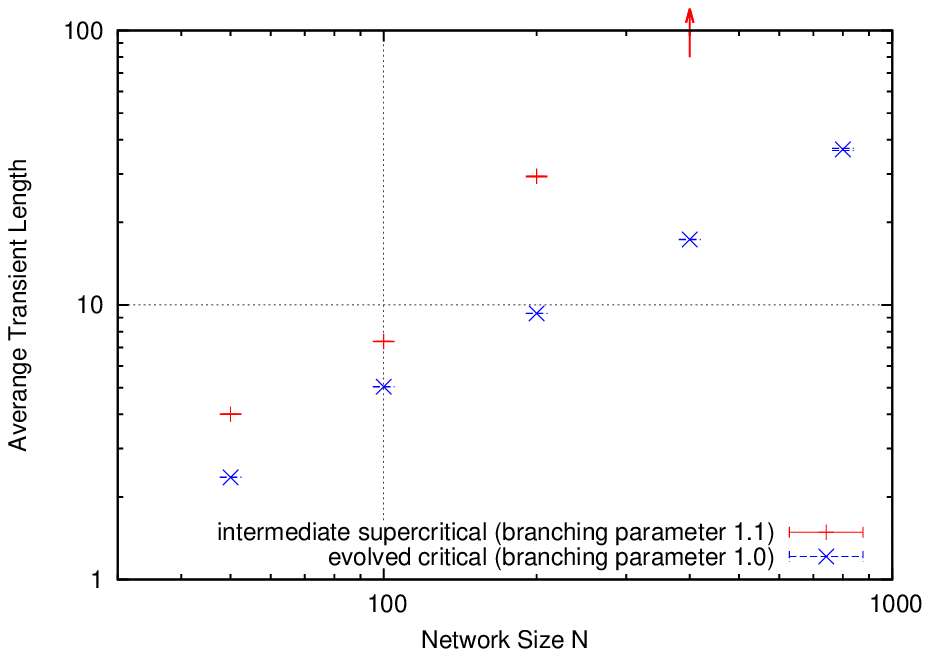}
	\caption{\small Left: cumulative distribution of avalanche sizes in an evolved sample network of $N=1024$ nodes. We find a broad, near power-law distribution comparable to a slope of $-1/2$, indicative of a critical branching process and corresponding nicely to the experimental results of Beggs and Plenz. Right: scaling of average transient lengths at different network sizes, 50 evolved networks each.}
	\label{fig:01-avalanche-sizes-transient-scaling}
\end{figure}

In addition to the branching parameter measurement, we also take a look at some dynamical properties of the evolved networks to further verify their criticality. As stated in the introduction section, we are especially interested in the resulting activity avalanches on the networks.  Figure~\ref{fig:01-avalanche-sizes-transient-scaling} (left) shows the cumulative distribution of avalanche sizes in an evolved sample network of $N=1024$ nodes. We observe a broad distribution of avalanche sizes and a power-law like scaling of event sizes with an exponent close to $-1/2$, corresponding to an exponent of $-3/2$ in the probability density -- which is characteristic of a critical branching process. At the same time, this is in perfect agreement with the event size distribution observed by Beggs and Plenz in their \emph{in-vitro} experiments.

If we randomly activate small fractions of nodes in an otherwise silent network (single nodes up to $\approx$ 5 \% of the nodes) to set off avalanches, we can also see (Figure~\ref{fig:01-avalanche-sizes-transient-scaling}, right) that the resulting transient length shows a power-law scaling with network size right up to network snapshots taken after an evolution run at the final average branching parameter of one. Intermediate networks taken from within an evolution process at a higher branching parameter instead show a superpolynomial increase of transient lengths with system size, which is precisely what we expect.

\subsection{Response to external perturbations}

In recent \emph{in-vitro} experiments, \cite{Plenz:2012} 
could further demonstrate that cortical networks can self-regulate in response to external influences, retaining their functionality while avalanche-like dynamics persist -- for example after neuronal excitability has been decreased by adding an antagonist for fast glutamatergic synaptic transmission  to the cultures.

To reproduce such behavior in our model, we can include variations in the activation thresholds $\Theta_i$ of the individual nodes, which had been set to zero in the above discussions for maximum model simplicity. Assume we start our network evolution algorithm with a moderately connected, but subcritical network, where all nodes have a higher activation threshold of $\Theta_i = 1$. According to the update rule \eqref{eq:01-update}, now at least two positive inputs are necessary to activate a single node. As the rewiring algorithm is based on propagation of thermal noise signals, the inverse temperature $\beta$ needs to be selected at a lower value than before. (As a general rule, $\beta$ should be selected in a range where thermal activation of nodes occur at a low rate, such that avalanches can be triggered, but are not dominated by noise.)

\begin{figure}[h!tbp]
	\centering
	\includegraphics[width=85mm]{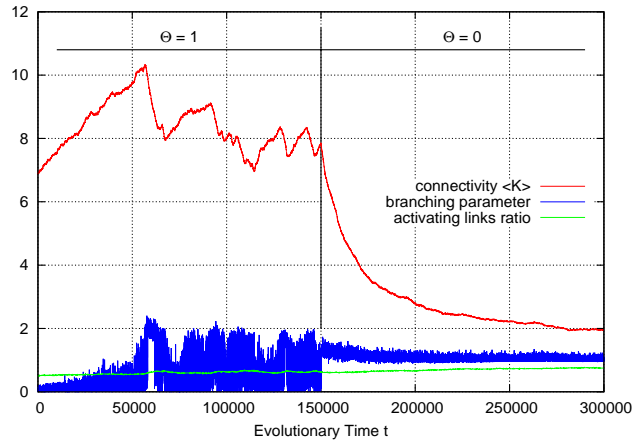}
	\caption{\small Rewiring response to a sudden decrease of activation thresholds. All $\Theta_i$ were set from 1 to 0 in the same time step.}
	\label{fig:01-variable-theta}
\end{figure}

Figure~\ref{fig:01-variable-theta} shows that, same as above, the subcritical network starts to grow new links, thereby increasing the average branching parameter. Again, this process is stopped after the supercritical regime is reached. While the system does not approach to a phase transition as nicely as shown above for activation thresholds of zero (in fact the branching fluctuates much more around the target value of one), the general tendency remains: the rewiring mechanism reacts properly as the network drifts too far off from criticality. At one time step in the center of Figure~\ref{fig:01-variable-theta}, we suddenly reset all nodes to an activation threshold of $\Theta_i = 0$, simulating the addition of a stimulant. As we can expect, this immediately puts the network into a supercritical, chaotic state. This is reflected by the branching parameter, which now constantly stays above one and does not fluctuate below anymore. It is clearly visible that the rewiring mechanism promptly reacts and drives back the connectivity, until the branching parameter slowly converges towards one again. A similar behavior is also found if thresholds $\Theta_i$ are not changed all at once, but gradually during network evolution.

\section{Conclusion}

We have seen that already very minimalistic binary neural network models are capable of self-organized critical behavior. While older models show some drawbacks regarding biological plausibility originating directly from their nature as spin networks, we subsequently presented a possible transition to a self-organized critical, randomly wired network of Boolean node states with emerging dynamical patterns, namely activity avalanches, reminiscent of activity modes as observed in real neuronal systems. This is possible via a simple, locally realizable, rewiring mechanism which uses activity correlation as its regulation criterion, thus retaining the biologically inspired rewiring basis from the spin version. While it is obvious that there are far more details involved in self-organization of real neuronal networks -- some of which are reflected in other existing models -- we here observe a fundamental organization mechanism leading to a critical system that exhibits statistical properties pointing to a larger class of universality, regardless of the details of a specific biological implementation.

\paragraph*{Acknowledgements}
For the presentation of the earlier work from our group
(Sections \ref{sec:BornholdtRohlf} and \ref{sec:BornholdtRoehl}),
excerpts from text and figures of the original
publications \citep{Bornholdt:2000a,Bornholdt:2003} were used, as
well as from \citet{Rybarsch:2012a,Rybarsch:2012b} for our recent work.

\bibliographystyle{apalike}
\bibliography{PlenzBook}

\end{document}